\documentclass{elsart}

\usepackage{psfig}

\usepackage{natbib}

\begin{document}



\begin{frontmatter}

\title{Optical, UV, and X-ray Clues to the Nature of Narrow Line AGNs}

\author[Tech]{Ari Laor}
\address[Tech]{Physics Department, Technion - 
Israel Institute of Technology, Haifa 32000, ISRAEL}

\begin{abstract}
AGNs with narrow Balmer lines show various extreme properties. 
In particular, rapid X-ray variability, steep X-ray spectra, 
peculiar optical and UV line ratios, and possibly peculiar line profiles.
Since all these phenomena occur together they are likely to be related
to one specific underlying physical parameter. I review recent
evidence, based on HST imaging of low $z$ quasars, which suggests that the 
H$\beta$ line width and continuum
luminosity of quasars provide a reasonably accurate estimate of the black
hole mass. This implies that narrow-line AGN have relatively 
low black hole masses, and thus high $L/L_{\rm Edd}$, as independently suggested
based on their steep X-ray spectra. I present additional
evidence suggesting that the X-ray variability and the radio loudness are
primarily driven by the black hole mass.
The high mass inflow rate into the core of narrow-line AGNs may produce a denser 
and more enriched BLR,
a high column radiation pressure driven outflow, and a smaller illumination
angle for the NLR, as suggested by the observed emission
line properties. Narrow-line AGNs may thus provide important clues for understanding
the rich overall phenomenology of AGNs.
\end{abstract}


\end{frontmatter}


\section{The Soft X-ray Clues}

The term `Narrow-Line Seyfert 1 Galaxies' (NLS1s) was coined by
Osterbrock \& Pogge (1985) who noted the overall peculiar optical 
emission-line spectra of Seyfert galaxies with narrow Balmer lines
(see Pogge, these proceedings). Follow-up studies of their radio emission
and optical polarization properties did not reveal anything
outstanding. The first hint for their remarkable X-ray properties
was found by Stephens (1989) who noted, based on {\em Einstein} data, 
that ``X-ray selection may be an 
efficient way to find NLS1s.'' This conclusion was much strengthened 
by Puchnarewicz et al. (1992) who found that $\sim 50$\% of their
{\em Einstein} ultrasoft survey AGNs were NLS1s, thus establishing that
NLS1s have steeper than usual soft X-ray spectra. This result was further
refined by Laor et al. (1994) who noted a remarkably strong correlation
between the H$\beta$ FWHM and the {\em ROSAT} $\alpha_x$ in a 
sample of 10 PG quasars ($r_s$ = 0.842, Pr=$2\times 10^{-3}$ ). 
This result was further strengthened when the
complete sample of all 23 $M_B<-23$, $z<0.4$, 
$N_{\rm H~I}<1.9\times 10^{20}$~cm$^{-2}$,
PG quasars was analyzed (Laor et al. 1997a, hereafter L97; 
$r_s$= 0.79, 
Pr= $7\times 10^{-6}$). Boller, Brandt \& Fink (1996) studied a large
sample of NLS1s with {\em ROSAT} and noted the clear absence of broad line 
AGNs with a steep $\alpha_x$. However, they found a much larger scatter in
the H$\beta$ FWHM vs. $\alpha_x$ relation. In particular, some of their NLS1s
show normal $\alpha_x$ values, unlike the sample of L97 where
all narrow-line quasars display steep $\alpha_x$ values. This 
difference most likely results
from the large luminosity range of the AGNs in the Boller et al.
sample. In particular, a plot of $\alpha_x$ vs. $L_X$ for the Boller et al.
sample (using the data in their Table 1) reveals a clear trend of flattening 
of $\alpha_x$ with decreasing $L_X$. All the flat ($\alpha_x>-2$)
NLS1s in the Boller et al. sample have low luminosity 
($L_X<1.3\times 10^{44}$~erg~s$^{-1}$), and all their luminous AGNs 
($L_X>1.3\times 10^{44}$~erg~s$^{-1}$) are steep ($\alpha_x<-2$). Thus, the
Boller et al. sample indicates that a significant H$\beta$ FWHM vs. $\alpha_x$
correlation appears in bright AGNs, and not when lower luminosity Seyferts
are included,
consistent with the strong correlation in the Laor et al. sample 
which includes only $M_B<-23$ PG quasars.
The H$\beta$ FWHM vs. $\alpha_x$ correlation thus involves luminosity
as well. The luminosity dependence can be understood if the primary driver of this 
correlation is $L/L_{\rm Edd}$, rather than just the H$\beta$ FWHM (see \S 2).
It will be interesting to explore if lower luminosity AGNs do
follow an H$\beta$ FWHM vs. $\alpha_x$ correlation, but offset towards flatter
$\alpha_x$.

\section{What Underlies the $\alpha_x$ vs. H$\beta$ FWHM Correlation?}

The $\alpha_x$ vs. H$\beta$ FWHM correlation is remarkably strong. 
Excluding luminosity luminosity correlations, it is the strongest of the 
294 different correlations measured by L97 (\S 3.1 there). Why
should the 0.2--2~keV continuum slope, most likely  
generated by the inner accretion disk ($R\sim 2-10R_g$), have anything 
to do with the width of the Balmer lines produced in the Broad Line Region
(BLR, $R\sim 10^4R_g$)?  The strength of this correlation suggests that
both parameters are controlled mostly by a single physical parameter. What is this 
parameter?

First, what is the width of the Balmer lines telling us? 
As discussed in various
papers (e.g. Puchnarewicz et al. 1992; Laor et al. 1994; Boller et al. 1996), 
there are three plausible 
explanations: 1. The BLR velocity distribution is anisotropic, and in
NLS1s we have a face-on view of a flattened BLR. 2. The distance of the BLR 
from the center is non-uniform, and in NLS1s it may be larger than usual. 
3. The black hole mass is non-uniform, and in NLS1s it may be 
lower than usual. 

What is the steepness of $\alpha_x$ telling us?  White, Fabian \& Mushotzky
(1984), and more recently Pounds, Done \& Osborne (1995), 
noted a possible analogy of AGNs to Galactic Black Hole candidates, where 
$\alpha_x$ becomes steeper 
in their `high state.' Pounds et al. therefore suggested that 
NLS1s, are `high state,' or high $L/L_{\rm Edd}$ AGNs. 
This suggestion basically overlaps explanation \# 3 above for the narrow H$\beta$.
A narrow H$\beta$ implies a low $M_{\rm BH}$, therefore a high $L/L_{\rm Edd}$, and 
therefore a steep $\alpha_x$.

Why should the H$\beta$ FWHM be so strongly tied to $M_{\rm BH}$?  This can
be understood with the following two 
assumptions: 1. The BLR velocity field is dominated
by gravity, i.e. $\Delta v^2\simeq GM_{\rm BH}/R_{\rm BLR}$
(see recent evidence in Peterson \& Wandel 1999).
2. The size of the BLR is set by the bolometric luminosity,
specifically $R_{\rm BLR}=0.1L_{46}^{1/2}$~pc, where $L\equiv L_{\rm Bol}$.  
The $L^{1/2}$ dependence is
indirectly inferred by the weak, if any, dependence of the BLR clouds' density
and ionization on luminosity, and it is expected theoretically if the gas in AGNs
is dusty (Netzer \& Laor 1993). It is experimentally
verified in reverberation mappings of AGNs (Kaspi et al. 2000, though apparently 
with a somewhat steeper slope of $\sim 0.7$).
Combining the two expressions for $M_{\rm BH}$ and $R_{\rm BLR}$ gives 
$m_9=0.18\Delta v^2_{3000} L_{46}^{1/2}$, where $M_{\rm BH}=10^9m_9M_{\odot}$,
and therefore $L/L_{\rm Edd}=0.44\Delta v^{-2}_{3000} L_{46}^{1/2}$
(Laor 1998). Luminous AGNs with 
narrow H$\beta$ thus have particularly high $L/L_{\rm Edd}$,
but low luminosity Seyferts with a similar H$\beta$ width, will have a lower
$L/L_{\rm Edd}$. This may explain why low luminosity NLS1s in the Boller
et al. sample do not have a steep $\alpha_x$, they may simply not have
a high $L/L_{\rm Edd}$.

The $L/L_{\rm Edd}$ interpretation thus seems to provide a very appealing
explanation for the H$\beta$ FWHM vs. $\alpha_x$ correlation.
But, how reliable is the $M_{\rm BH}(\Delta v, L)$ determination?
This mass estimate has been around for many years (e.g. Dibai 1981;
Wandel \& Yahil 1985; Padovani \& Rafanelli 1988; Peterson et~al. 1998), but 
it was generally viewed as highly uncertain. Possible problems include: 
1. $R_{\rm BLR}(L)$ is determined through reverberation mostly in AGNs
with ``normal'' $\Delta v$. Do NLS1s simply have a larger 
$R_{\rm BLR}(L)$?
2. $L$ may be anisotropic. Is $R_{\rm BLR}$ in NLS1s larger because the
BLR sees a brighter ionizing continuum than we do?
3. $v$ may be anisotropic. Is the BLR in NLS1s simply seen `face-on'?
4. $v$ may be dominated by radiation
or magnetic pressure, and thus not related to $M_{\rm BH}$.

Given this host of possible systematic effects, one cannot attach a reliable 
error bar to $M_{\rm BH}(\Delta v, L)$ (though recent reverberation 
studies suggest that problems \#1 and \#4 may not be significant). 
Below, I describe recent new evidence that provides
an indirect check on the $M_{\rm BH}(\Delta v, L)$ estimate, and 
indicates it is most likely accurate to within a factor of 2--3.

\section{An Independent Check of the $M_{\rm BH}(\Delta v, L)$ Estimate.}

Magorrian et al. (1998, and references therein) have recently obtained two 
outstanding results:
1. Most (possibly all) early type galaxies host a Massive Dark Object (MDO)
at their center, most likely a massive black hole. 2. The MDO mass is
correlated with the bulge mass. These results, if true, have a fundamental
impact on our understanding of quasar and galaxy
formation and evolution (e.g. Kauffmann \& Haehnelt 2000).

The Magorrian et al. study is based on HST observations of an unbiased sample 
of 36 nearby early type galaxies (i.e. their selection was independent of a~priori
knowledge about core kinematics). They combined ground based spectroscopy 
with HST imaging and constructed
simplified kinematic models with a constant $M/L$ plus a spatially unresolved MDO. 
Their simplified model allowed an acceptable fit in 32 galaxies. In 31 of these
galaxies a MDO is allowed (best fit $M_{\rm MDO}>0$), and in 26 it is required
(at $>95$\% significance level). The $M_{\rm MDO}$ vs. 
$M_{\rm bulge}$ correlation found in this study cannot be
a selection effect since 31/32 MDOs were detected. A study which is based on  
a compilation of all published results would be strongly biased, as non detections
are generally not published, and the minimum detectable $M_{\rm MDO}$
is strongly correlated with $M_{\rm bulge}$. This bias is not present in the 
Magorrian et al. study, and unless their $M_{\rm MDO}$ values are wrong, the 
correlation they find is highly significant. Ongoing spectroscopy with HST 
should significantly improve the kinematic constraints on $M_{\rm MDO}$.

How is all of the above related to quasars?  The first hint came from the study 
of McLeod \& Rieke (1995), who found that there is an upper limit
for the quasar luminosity at a given host luminosity, a limit which grows
roughly linearly with the host luminosity. McLeod (1997) speculated that 
this limit may be $L_{\rm Edd}$, if
$M_{\rm BH}$ has an upper limit in a host of a given luminosity, and if this
limit increases with the host luminosity. 

Do quasar hosts follow the $M_{\rm BH}$ vs. $M_{\rm bulge}$ correlation 
suggested for normal galaxies?  
To explore this, one needs a high-quality sample of quasars (uniform, well defined, 
complete), with a high quality data set on their host luminosity. The closest
we could get to this ideal is with the PG sample. Bahcall et al. (1997) and
Kirhakos et al. (1999) provide
careful and systematic measurements of the host luminosities of 23 quasars
imaged with HST, of
which 15 are PGs. Boroson \& Green (1992, BG92) gives high quality 
spectroscopy, and Neugebauer et al. (1987) provide well calibrated 
spectrophotometry.
Combining $L$ and $\Delta v$ provides
$M_{\rm BH}(\Delta v, L)$. Figure 1 presents the bulge luminosity
vs. $M_{\rm BH}$ relation for these 15 PG quasars, together with the 
distribution of the nearby
galaxies from Magorrian et al.
The quasar correlation is highly 
significant ($1.74\times 10^{-3}$); it also overlaps
the distribution of nearby galaxies, and it indicates a non-linear
$M_{\rm BH}$ vs. $M_{\rm bulge}$ relation. A least-squares fit to the quasars 
gives $M_v$(bulge)$=-21.76-(1.50\pm 0.38)\log m_9$, which implies
$M_{\rm BH}\propto M_{\rm bulge}^{1.4}$, unlike the linear relation proposed
by Magorrian et al.
It is interesting to note that the three most accurate $M_{\rm BH}$ determinations,
M~87, NGC~4258, and the Galaxy, agree well with the non-linear quasar relation,
as does the only NLS1 in the Bahcall et al. sample, NAB~0205+024.

\begin{figure}[htb]
\centerline{\psfig{figure=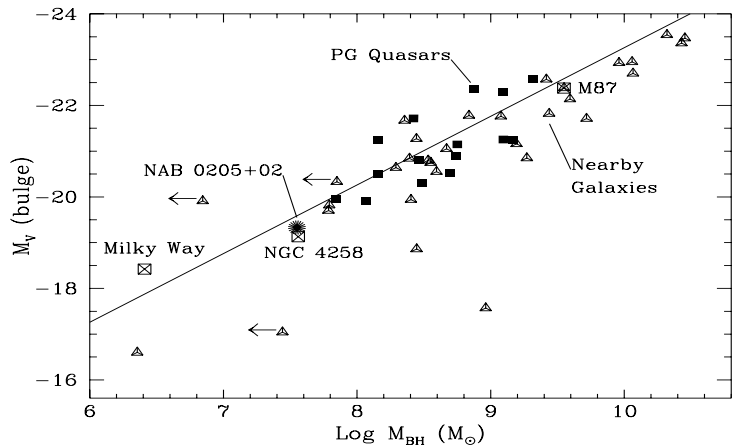,height=2.4truein,width=4.truein,angle=270}}
\caption{The bulge luminosity vs. $M_{\rm BH}$ relation for 
nearby normal
galaxies (Magorrian et al.) and for PG quasars (Laor 1998). The positions of the
three most accurate $M_{\rm BH}$ determinations are indicated. NAB~0205+024 is a 
NLS1 from Bahcall et al.}
\end{figure}

The overlap of the PG quasar and the Magorrian et al. galaxy distributions
in Fig.1 is remarkable given the fact that these are {\em apparently unrelated} 
types of objects, and that one is using {\em completely different} methods to 
measure $M_{\rm BH}$. This overlap suggests that the 
$M_{\rm BH}(\Delta v,L)$ estimate
in quasars is probably accurate 
to a factor $\sim 2-3$, and provides the first indirect check
for $M_{\rm BH}(\Delta v,L)$. Is $M_{\rm BH}$ directly related to some
of the peculiar emission properties
of NLS1s?

\section{X-ray Variability and $M_{\rm BH}$.}

One of the outstanding properties of many NLS1s is their rapid and large amplitude
soft X-ray variability (e.g. Boller et~al. 1996 and references therein).
More systematic results came from the pilot study of Fiore et al. (1998),
who observed a complete and unbiased small sample of six PG quasars which included the
three narrowest and three broadest H$\beta$ quasars in the L97 sample. All
the steep quasars varied significantly on a $10^5$--$10^6$~s timescale, while none
of the flat $\alpha_x$ quasars did (both groups varied similarly on the $10^7$~s
timescale), hinting at a strong correlation between 
X-ray variability and H$\beta$ FWHM.

The variability of a much larger and heterogeneous sample of AGNs 
was explored by Turner et al. (1999) on a $10^4$~s timescale. 
They recovered the well known  
variability vs. luminosity correlation, but discovered there is a much tighter 
correlation of the variability amplitude with the H$\beta$ FWHM. A qualitatively
similar result was obtained by Leighly (1999), who found that NLS1s obey the
variability amplitude vs. $L$ relation of broad-line AGNs, but were displaced
upwards towards larger variability amplitudes at a given $L$. What is the underlying
physical parameter which controls these trends?  Is it $M_{\rm BH}$, or possibly 
$L/L_{\rm Edd}$?

Comparison with GBHCs provides some clues.
GBHCs can vary significantly down to ms timescales, and most likely harbor a few 
$M_{\odot}$ black holes. Seyfert galaxies can vary significantly down to ks timescale, 
and most likely
harbour $\sim 10^7-10^8~M_{\odot}$ black holes. Based on this simple empirical fact, it
appears plausible that the variability timescale (at a fixed variability
amplitude) is roughly proportional to $M_{\rm BH}$. This can occur if flux
modifying disturbances have some characteristic velocity in the X-ray emitting region 
(independent of absolute luminosity),
and if the size of this region scales with mass (i.e. it is fixed in units of
$R_g=GM/c^2$). 

Turner et al. provide $L_x$ and H$\beta$ FWHM for all their objects. I used these
parameters to obtain a rough estimate of $M_{\rm BH}$ and $L/L_{\rm Edd}$.
For objects with more than one value of $\sigma^2_{\rm RMS}$ in the Turner et al. 
paper, I used a mean value (this reduces the scatter).
Figure~2 presents plots of the correlations of $\sigma^2_{\rm RMS}$ with
$L_x$, H$\beta$ FWHM, and with the combinations roughly proportional to $M_{\rm BH}$
and $L/L_{\rm Edd}$ (Spearman $r$ and probabilities are indicated in each panel).
The correlation of $\sigma^2_{\rm RMS}$ with $M_{\rm BH}$ is the strongest
(significance level $5.5\times 10^{-9}$), which suggests that  $M_{\rm BH}$
is the underlying physical parameter which drives the observed dependence
of the X-ray variability on luminosity and on line width. AGNs may have
a universal power spectrum density (PSD) of X-ray fluctuations, and 
$M_{\rm BH}$ may just set the timescale at a given amplitude 
(Hayashida, these proceedings).
Interestingly, there
appears to be a certain $L/L_{\rm Edd}$ above which the mean
$\sigma^2_{\rm RMS}$ increases by a facor of $\sim 10$. This may reflect
a qualitative change in the nature of the X-ray variability. Does the inner, 
X-ray emitting, accretion disk 
become strongly unstable above a certain fraction of $L_{\rm Edd}$?

\begin{figure}[htb]
\centerline{\psfig{figure=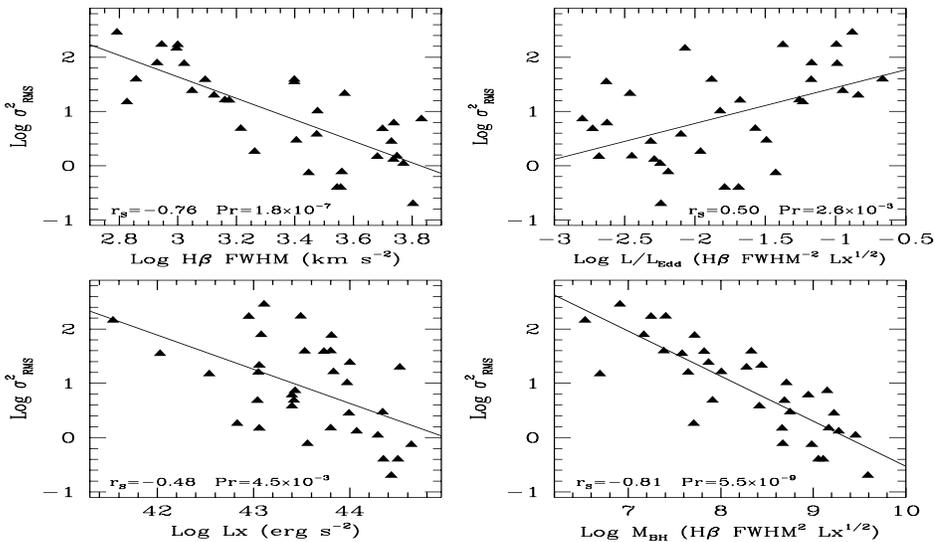,height=3.truein,width=5.truein,angle=0}}
\caption{The normalized RMS X-ray variability, $\sigma_{\rm RMS}$,
of 36 AGNs from Turner et al. (1999). Left panels show 
$\sigma_{\rm RMS}$ vs. $L_x$ and H$\beta$ FWHM.
Right panels show $\sigma_{\rm RMS}$ vs. the combinations which are approximately
proportional to $L/L_{\rm Edd}$ and $M_{\rm BH}$.}
\end{figure}

Leighly (1999) provides similar data for a partly overlapping sample which
includes only NLS1s.
This sample shows significantly weaker correlations, which may be
partly due to the smaller ranges in H$\beta$ FWHM and $L$ covered by that sample.
A proper study of the above correlations requires a much more systematic study
of a well defined sample, including accurate bolometric luminosities, and 
high-quality optical spectroscopy. Particularly careful analysis is required for 
low luminosity AGNs where NLR and host galaxy contaminations can be significant. 

\section{Radio Loudness and $M_{\rm BH}$.}

NLS1s are, with very few exceptions, radio quiet. The few radio-loud
NLS1s are only marginally loud (Siebert et al. 1999), 
and may be intrinsically radio-quiet quasars where a weak jet is 
beamed at us. Why are there no radio-loud NLS1s?  In fact, why are
there no radio-loud Seyferts? Recent studies with HST have clearly demonstrated
that all radio-loud quasars reside in elliptical (or interacting) 
galaxies, and that all quasars with spiral hosts are radio quiet. 
Being radio loud means having a powerful jet. The jet
is formed within a mpc of the center, where the massive black hole resides. 
The fundamental puzzle is {\em how does the inner mpc know about the 
host type?}  The answer may lie in the $M_{\rm BH}$ vs. $M_{\rm bulge}$
correlation. The jet formation is set by the black hole, and the black hole
properties are related to the bulge properties. 
Spirals have 
small bulges, thus small black holes, and these do not produce powerful jets.
On the other hand, bright ellipticals generally host very massive black holes,
and these always channel much of the accreting gas into powerful relativistic 
outflows. Is the production of powerful jets directly
related to the black hole mass? 

\begin{figure}[htb]
\centerline{\psfig{figure=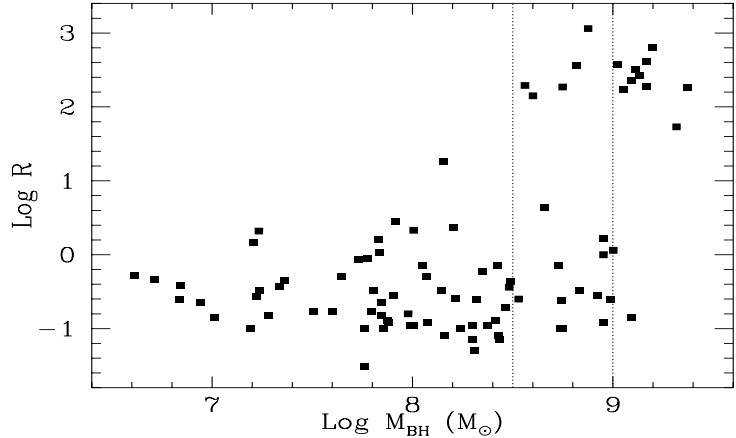,height=2.4truein,width=4.truein,angle=270}}
\caption{Radio loudness $R$ vs. $M_{\rm BH}$ for the 87 PG Quasars in BG.
Practically all $M_{\rm BH}<3\times 10^8M_{\odot}$ quasars are radio quiet ($R<1$),
and nearly all $M_{\rm BH}>10^9M_{\odot}$ quasars are radio loud ($R>1$) quasars.}
\end{figure}

This question can be explored directly, without using any information about 
the host properties. Figure 3 shows a plot of $R$ vs. $M_{\rm BH}(L, \Delta v)$
for all 87 PG quasars from the BG sample,
where $R$ 
($\equiv f_{\rm 5~GHz}/f_{\rm opt}$) is taken from Kellerman et~al. (1989). The 
answer appears to be positive. All quasars with $M_{\rm BH}>10^9M_{\odot}$
are radio loud, and practically all quasars with 
$M_{\rm BH}<3\times 10^8M_{\odot}$ are radio quiet. This then provides a 
phenomenological understanding of  
why Seyferts (inc. NLS1s), which do not have very massive bulges, are practically 
always radio quiet, and why radio-loud quasars require hosts with massive
bulges. The remaining puzzle is {\em why should the formation of a powerful jet be so 
critically dependent on $M_{\rm BH}$?}

\section{Clues from the UV Emission Lines \& Other Clues}
Careful systematic studies of the UV emission line properties of
complete samples of radio-quiet AGNs are beginning to emerge, and these clearly
indicate that narrow-line AGNs have characteristic spectra, which are different
from those of normal AGNs (e.g. Wills, these proceedings). The best studied NLS1
galaxy in the UV is the prototype, I~Zw~1 (Laor et al. 1997b). Below I'll 
briefly describe some of our results and what they may imply. 

I~Zw~1 shows unusual emission line properties. In particular, very weak C~III], 
strong Al~III, strong Fe~III, and generally strong low ionization lines.
Wills et al. (1999) explored a sample of 22 PG quasars, and found that the
properties of I~Zw~1 are common among NLS1s. In addition they found
that many of the UV emission line properties show strong trends with the 
H$\beta$~FWHM. The observed trends probably reflect an increase in the density
and metalicity of the BLR as the H$\beta$~FWHM decreases.

I~Zw~1 also displays interesting trends among its UV emission line profiles. 
The line peaks get progressively more
blueshifted as the ionization level increases, rising from zero shift for O~I, Mg~II,
and Si~II, to 250--500~km~s$^{-1}$ for Ly$\alpha$, Si~III], C~III], and Al~III, to 
$\sim 900$~km~s$^{-1}$ for C~IV and N~V, and finally 
to $\sim 2000$~km~s$^{-1}$ for He~II.
This trend is seen in other AGNs as well, but the amplitude of the blueshifts in
I~Zw~1 is about 4 times larger than in typical AGNs. In addition, the UV lines
develop increasing excess blue wing flux with increasing ionization. These
trends may be interpreted as an outflowing component in the BLR, which is seen
on the approaching side, but is obscured on the receding side (due to a
disk?). The outflowing gas needs to get progressively more highly ionized as its
velocity increases. We may actually be observing the edge of this outflow
in I~Zw~1,
as it shows a weak absorption system in Ly$\alpha$, N~V, C~IV, and Si~IV 
with a blueshift of 1870~km~s$^{-1}$. 

Is such an outflowing BLR component typical of NLS1s?  We do not know yet,
but one can speculate as to why it may be common. Bright NLS1s  have a
high $L/L_{\rm Edd}$. Radiation 
pressure which is incident on the BLR clouds must ablate a surface layer
and drive an outflow. The radiation pressure can overcome gravity 
in a surface layer with a column density of 
$N_{\rm H}\le 1.5\times 10^{24}L/L_{\rm Edd}$~cm$^{-2}$. Thus, the
outflows in high $L/L_{\rm Edd}$ AGNs will be relatively thick, and
may have enough emission measure to produce a noticeable contribution to
the UV emission lines which comes from the denser cores of the BLR clouds. As the
velocity of the outflow increases, its density drops (due to mass
continuity), and its ionization state increases, thus explaining the 
observed profile trends with ionization level.

The UV and X-ray absorption properties of NLS1s may also be different
from those of normal AGNs, as further discussed in Brandt et al. (2000), and
Brandt (these proceedings). 
NLS1s also show peculiar optical emission line ratios, in particular strong
Fe~II, and weak [O III], as is clearly established in the seminal BG paper.
These correlations were in fact already noted by Boroson \& Oke (1984) and
Boroson, Persson, \& Oke (1985), who found that the above parameters also 
correlate with the host galaxy colors, the equivalent width of the 
{\em extended} [O III] emission, and the radio morphology. Amazingly enough, 
Boroson et al. suggested already
back then that the underlying physical parameter was $L/L_{\rm Edd}$!
This suggestion was based on the notion that the weakness of [O III] 
(extended and nuclear) is due to the thicker accretion disk which collimates
the ionizing continuum over a smaller solid angle, and reduces the NLR
illumination. It is remarkable that this suggestion was again reached based
on each of the following unrelated properties; rapid X-ray variability, 
steep X-ray slope, and the narrow H$\beta$.

\section{Should the Definition of NLS1s be Revised?}

Yes, in part.
The current working definition for NLS1s is [O~III]/H$\beta<3$ and
H$\beta$~FWHM$\le 2000$~km~s$^{-1}$. The first part is well justified. 
A given solid angle of the NLR typically produces equivalent widths 
[O~III]:H$\beta\simeq 10:1$, while the same solid angle at the BLR produces
about [O~III]:H$\beta\simeq 0:1$. Thus, 
[O~III]:H$\beta\simeq \Omega_{\rm BLR}\times 0:1+\Omega_{\rm NLR}
\times 10:1<3 \rightarrow \Omega_{\rm BLR}/\Omega_{\rm NLR}>2$, 
i.e. it ensures that $>2/3$ of H$\beta$ 
is from the BLR. The narrowness of H$\beta$ is thus most likely not just an NLR
contamination effect (though it's always better to verify that by subtracting
a properly scaled [O~III] profile from H$\beta$).

What about the H$\beta$~FWHM criterion?  The observed distribution of
H$\beta$~FWHM is not bimodal, and thus there is no natural H$\beta$~FWHM 
cutoff for
NLS1s (in contrast with, e.g., $R$ of radio loudness). Furthermore, 
it is now clear
that NLS1s lie at the extreme end of a continuous distribution of emission
properties, and therefore the specific H$\beta$~FWHM cutoff is obviously
arbitrary. The crucial question is what underlying extreme physical parameter
are we interested in?  If we are looking for high $L/L_{\rm Edd}$ objects,
say $L/L_{\rm Edd}>1$, that translates (see \S 2) to 
$\Delta v<2000 L_{46}^{1/4}$~km~s$^{-1}$. Thus, the H$\beta$~FWHM cutoff needs to
be luminosity dependent.\footnote{The BG database suggests that 
$\Delta v_{\rm min}\sim 2000 L_{46}^{1/4}$~km~s$^{-1}$, i.e.
$L/L_{\rm Edd}\le 1$} For example, 
an $M_V=-27$
quasar with H$\beta$~FWHM$\sim 2000$~km~s$^{-1}$ should have about the same 
$L/L_{\rm Edd}$ as
an $M_V=-22$ Seyfert with H$\beta$~FWHM$\sim 600$~km~s$^{-1}$. 
If one is looking for low $M_{\rm BH}$ objects (e.g. for rapid X-ray variability),
say $M_{\rm BH}<10^7M_{\odot}$, that translates to 
$\Delta v<700 L_{46}^{-1/4}$~km~s$^{-1}$, i.e. one can accommodate broader
lines as one goes to lower $L$.

\section{Conclusions, Speculations and Open Questions}

The highly simplified $M_{\rm BH}(\Delta v, L)$ estimate in quasars appears
to be accurate to within a factor of about 2--3. NLS1s thus most likely have 
a relatively low 
$M_{\rm BH}$, and if they are not very faint, a relatively high $L/L_{\rm Edd}$.
The rapid X-ray variability of NLS1s is mostly due to their low $M_{\rm BH}$,
but there may also be some enhancement of variability for NLS1s with the highest
$L/L_{\rm Edd}$. NLS1s, and Seyferts in general, are radio quiet most likely because
of their relatively low $M_{\rm BH}$. The UV lines suggest the BLR is denser and
possibly more enriched. A possible scenario is that the high $L/L_{\rm Edd}$
results from large amounts of gas being dumped into the center of the galaxy. This
increased accretion rate brings in denser gas (implied by the UV line ratios), enhances 
star formation rate and therefore metalicity (implied by the strong N~V), and possibly 
blocks most of the NLR illumination (implied by the weak narrow lines). The increased
$L/L_{\rm Edd}$ may enhance the column of the radiation-pressure ablated surface
layer in the BLR (implied by the UV line profiles).

Clearly, there are many open questions which need to be addressed
to reliably establish or disprove the above scenario. 
Specifically: 1. Do NLS1s follow the $M_{\rm BH}$ vs. $L_{\rm bulge}$ relation?
The best way to proceed here is to extend the analysis of Bahcall et al.
and Kirhakos et al. to lower luminosity PG Quasars which qualify
as NLS1s. 2. How tight is the X-ray variability vs. $M_{\rm BH}$ relation?
One needs a well defined sample, high quality optical spectroscopy, and long
X-ray integrations (if the PSD is non-stationary).
3. Do NLS1s generally show the UV emission-line profile trends seen in I~Zw~1?
One needs high-quality UV spectroscopy for a well defined sample of NLS1s,
and the PG sample can again be a very useful parent sample. 4. Are AGN absorption
outflows related to $L/L_{\rm Edd}$?  The ideal route here is to survey the
UV absorption properties of AGNs spanning a large range in $L$ and in
$L/L_{\rm Edd}$ (the PG sample again!). 5. What controls
the strength of [O III]? Is it the NLR gas covering factor? Ionization state?
Density? A careful study of various line ratios in the NLR is required for
a conclusive answer. 6. What controls the other intriguing correlations noted by 
Boroson et al (host colors, extended [O III], and radio morphology)?
Follow-up studies of these correlations with higher quality data is the
first step required here.

I thank the organizing committee for their kind invitation, generous
support, and for a very interesting meeting. 



\end{document}